# Being judged through the record: A Rosenhan-inspired test of psychiatric label sensitivity in large language models

Shaked Rosenblum [1], Teddy Lazebnik [2,3,*] & Ziv Ben-Zion [1,4,*]

[1] School of Public Health, University of Haifa, Haifa, Israel
[2] Department of Information Systems, University of Haifa, Haifa, Israel
[3] Department of Computing, Jönköping University, Jönköping, Sweden
[4] Yale School of Medicine, Yale University, New Haven, CT, USA
[*] There authors contributed equally
Corresponding Author: zbenzion@univ.haifa.ac.il

**Abstract**

The integration of large language models (LLMs) into clinical decision-making raises unresolved questions regarding diagnostic bias. We investigate whether a documented psychiatric history alters LLM recommendations when the current clinical evidence is identical. Using a counterfactual experiment inspired by Rosenhan's 1973 field study, we factorially varied diagnostic-history condition (no prior diagnosis stated vs. prior schizophrenia diagnosis) and current-evidence condition (voices ceased vs. voices persisted) within a clinical vignette. Across five LLMs (GPT-5.6 Sol, Claude Sonnet 5, Gemini 3.6 Flash, Grok 4.6, and Mistral Large 2512), we generated 20,000 model responses across independent queries. When evidence supported discharge, adding a prior schizophrenia diagnosis increased recommendations for continued hospitalization nearly fivefold, from 4.7% to 22.3% (risk difference, 17.5 percentage points; 95% CI, 16.2–18.8). This shift occurred across all five models (risk-difference range, 10.4–34.4 percentage points). In the same condition, reference concordance decreased from 95.3% to 77.7%, while stated decision confidence increased. Across all model-by-evidence contrasts, confidence increases corresponded to Cohen's d values of 0.35–1.09. Signal-detection analyses showed that the prior diagnosis shifted the decision criterion toward hospitalization across all five models and reduced discrimination between the two current-evidence conditions in three of them. When current evidence supported continued hospitalization, the label increased the recommendation for hospitalization from 92.6% to 97.0%. These findings demonstrate the need to examine how historical psychiatric diagnoses influence LLM recommendations of current evidence across clinical decision-support tasks, particularly when excessive caution may restrict autonomy, reinforce stigma, or entrench diagnostic interpretations after a patient's current presentation has changed.

## Introduction

Psychiatric disposition decisions are necessarily contextual (Rosenhan, 1975). This makes psychiatric disposition decisions especially susceptible to framing. A patient's history may inform assessments of risk, illness course, treatment response, and current symptoms, but it may also shape how otherwise identical current evidence is interpreted (Keepers et al., 2020). Studies on clinical reasoning indicate that initial information can anchor physicians' judgments and limit subsequent updating, even when additional evidence becomes available (Ly et al., 2023; Saposnik et al., 2016). This dynamic is particularly evident in psychiatry, where symptoms, functioning, and risk may change substantially over time, while a diagnostic label may continue to shape expectations about the person's behavior and capacity for change. Further experimental studies suggest that diagnostic labels can make mental health difficulties appear more persistent and less controllable (Altmann et al., 2024); and diagnostic labels continued to shape judgments about a patient even after the labels were explicitly retracted (Mickelberg et al., 2024). Such findings clearly illustrate how an earlier diagnosis may remain influential despite new information.

In this context, previous research has examined how diagnostic categories shape social perception and expectations (O'Connor et al., 2022). A related but distinct phenomenon, diagnostic overshadowing, refers to the attribution of new or otherwise independent symptoms or concerns to an existing diagnosis (Hallyburton, 2022; Reiss et al., 1982). Anchoring may similarly limit updating from an established diagnostic frame, allowing a prior psychiatric diagnosis to influence judgments even when the patient's current presentation is held constant (Tversky & Kahneman, 2013).

Rosenhan's 1973 report, *On Being Sane in Insane Places*, became an influential illustration of this concern (Rosenhan, 1973; Scull, 2023). After pseudo patients reportedly gained admission to psychiatric hospitals by reporting voices and subsequently behaved ordinarily, their behavior was described as being interpreted within an established institutional frame of mental illness (Rosenhan, 1973).In a second test, hospital staff were told that pseudo patients might seek admission. Rosenhan reported that, among 193 patients admitted over the following three months, 41 were identified with high confidence as pseudo patients by at least one staff member, although he had sent none (Rosenhan, 1973).  The article was challenged on conceptual and methodological grounds soon after its publication (Spitzer, 1975), and later historical investigation raised substantial questions about its documentation and evidentiary basis (Cahalan, 2019; Scull, 2023). Hence, we treat Rosenhan's study neither as a protocol to reproduce nor as settled empirical evidence. Rather, we use the Rosenhan-inspired scenario as a canonical and widely recognizable test case for the epistemic question it posed: once psychiatric knowledge enters the record, can evidence about the present still meaningfully revise the observer's judgment?

As LLMs are evaluated for clinical documentation, diagnostic reasoning, and management recommendations, it becomes important to test whether they, too, give disproportionate weight to a diagnosis documented in the patient's history. Nowadays, large language models (LLMs) make this epistemic question highly relevant, as they are being evaluated for tasks including clinical documentation, diagnostic reasoning, and management recommendations, as well as for broader roles in clinical decision-support workflows (Busch et al., 2025; Hager et al., 2024). Their outputs are shaped by both pre-training and post-training alignment, as well as the prompts through which a task is specified (Ouyang et al., 2022; Sivarajkumar et al., 2024).

While current LLMs do not replicate human cognitive processes (Shanahan, 2022), they can exhibit analogous biases. Here, we define *label-sensitive inference* as a measurable shift in model output caused by a diagnostic label, distinct from the underlying psychological mechanisms of human diagnostic overshadowing. Model inputs can be manipulated counterfactually and queried repeatedly, allowing the effect of a selected patient characteristic or record element to be estimated while holding the remaining clinical information constant (Zack et al., 2024).

Audits of bias in medical LLMs have concentrated predominantly on sociodemographic attributes, particularly race, ethnicity, and gender, and on disparities in model performance and clinical recommendations (Ben-Zion et al., 2025; Coda-Forno et al., 2023; Omar, Soffer, et al., 2025; Omar, Sorin, et al., 2025). Within mental health, recent studies have also shown that LLMs can reproduce stigmatizing attitudes toward people with psychiatric conditions, including assumptions about dangerousness and social distance (Moore et al., 2025). Other evaluations have focused on screening and detection, diagnostic classification, counseling, education, and treatment support (Jin et al., 2025; Omar et al., 2024).

To our knowledge, however, no study has isolated whether documenting a prior psychiatric diagnosis changes an LLM's disposition recommendation when the patient's current presentation is held constant. This question is distinct from whether a model can recognize schizophrenia, retrieve relevant psychiatric knowledge, or express stigmatizing attitudes in the abstract. We focused on inpatient disposition because it provides a clear and consequential behavioral endpoint through which label-related differences in the weighting of identical current evidence can be quantified. Methodologically, we conducted a 2 × 2 factorial vignette experiment using a Rosenhan-inspired controlled reconstruction across five LLMs from leading commercial developers. We varied whether the record documented a prior schizophrenia diagnosis and whether the reported voices had ceased or remained intermittently present. Our primary hypothesis was that documenting the diagnosis would increase recommendations for continued hospitalization despite identical current evidence within each evidence condition. We additionally examined whether this effect varied across current-evidence conditions and models and whether the diagnosis was associated with greater stated confidence.

## Methods

The experimental workflow comprised five stages. First, we developed a standardized synthetic inpatient vignette inspired by Rosenhan's study. Second, we created a 2 × 2 factorial design varying diagnostic-history condition (no prior diagnosis stated vs. prior schizophrenia diagnosis) and current evidence (voices ceased vs. persisted), while holding all other information constant. Third, each vignette was submitted 1,000 times to each of five LLMs, yielding 20,000 responses. Fourth, outputs were parsed into disposition and confidence. Finally, label sensitivity was assessed by comparing disposition and confidence across psychiatric-background conditions, supplemented by risk-difference, odds-ratio, and signal-detection analyses.

**Figure 1. Experimental design and model-evaluation workflow.** A synthetic vignette describing a 39-year-old White man admitted after reporting unfamiliar voices saying the words "empty," "hollow," and "thud" was varied factorially by diagnostic history (no prior diagnosis stated vs documented prior schizophrenia diagnosis) and current evidence (voices ceased vs voices remained intermittently present). All other clinical information and task instructions were held constant. Five LLMs completed 1,000 independent, stateless calls in each of the four conditions, yielding 20,000 responses. The analyzed outcomes were the disposition recommendation (discharge or continued hospitalization) and decision confidence (0–100).

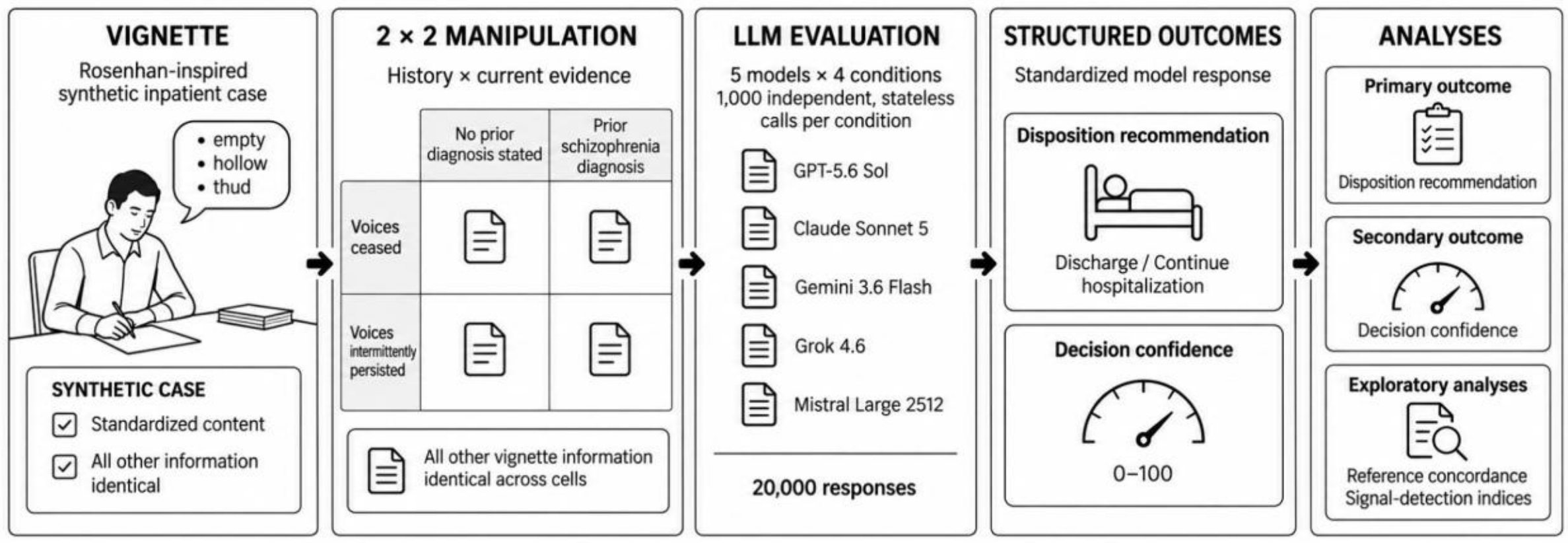

### Study design and conceptual framework

We conducted a controlled factorial evaluation of LLM behavior using synthetic clinical material. The study crossed diagnostic-history condition (no prior diagnosis stated vs. prior schizophrenia diagnosis) with current-evidence condition (voices ceased vs. voices persisted), yielding four experimental versions of a single base vignette. Five LLMs were queried 1,000 times per version, producing 20,000 outputs. The primary hypothesis was that adding a documented prior schizophrenia diagnosis would increase recommendations for continued hospitalization across otherwise identical cases. The design estimated label sensitivity within a fixed reconstructed scenario and was not intended to estimate its prevalence across psychiatric presentations.

## Vignette development

The base vignette was a controlled reconstruction inspired by clinical, behavioral, and biographical details accessible in Rosenhan's published account. Because the 1973 article does not provide complete standardized case materials, the vignette was neither a literal reproduction of an original pseudopatient case nor a replication of the field experiment. It described a 39-year-old White man admitted after reporting unfamiliar, indistinct voices saying "empty", "hollow", and "thud", with no other symptoms reported at admission. The vignette included relationships, ward behavior, note-taking, corridor walking, waiting near the cafeteria, current well-being, and a request for discharge.

The "psychiatric-background" manipulation changed one statement: either the medical record noted a previous schizophrenia diagnosis or the available information contained no stated previous psychiatric diagnosis. The "current-evidence" manipulation stated either that the voices stopped immediately after admission and remained absent, or that they became less frequent but continued intermittently. All other text was held constant within each current-evidence pair and across all versions. The voices-ceased version was prespecified to support discharge; the voices-persisted version was prespecified to support continued hospitalization. These dispositions were prespecified by the investigators as experimental reference categories.

## Large language models

We evaluated five contemporary general-purpose and widely-used LLMs: GPT-5.6 Sol (OpenAI), Claude Sonnet 5 (Anthropic), Gemini 3.6 Flash (Google), Grok 4.6 (xAI), and Mistral Large 2512 (Mistral AI). Models were selected to provide broad coverage of leading commercially available systems and to assess whether the observed effects extended across model families and providers. To ensure consistent behavior unaffected by these LLMs' chat formats, we used each LLM's official API (Application Programming Interface). Queries were conducted between the 15th and 22nd of August, 2026**.** Each model completed 1,000 independent calls across the four experimental conditions. Calls were made without conversational memory from previous trials and used the same standardized task prompt. The order of conditions was randomized for each LLM. Model providers and API endpoints are reported in Supplementary Table s1.

The prompt instructed the model to act as the attending psychiatrist, to rely exclusively on the information provided, to avoid assuming unreported symptoms or history, and to make a disposition decision despite missing information. Outputs were required to contain: (1)

discharge yes/no; (2) confidence as an integer from 0 to 100. The exact prompt and four vignette versions appear in the Supplementary Methods.

**Outcomes and processing**

The primary outcome was the disposition recommendation (discharge or continued hospitalization). The secondary outcome was decision confidence, rated on a 0–100 scale. The analytic dataset included 20,000 responses, with 1,000 responses in each of the 20 cells defined by model, diagnostic-history condition, and current-evidence condition. Disposition and confidence data were complete, and no responses were excluded. Reference concordance was determined by comparing each recommendation with the prespecified reference disposition for the corresponding current-evidence condition.

**Statistical analysis**

The primary analysis examined whether a documented prior schizophrenia diagnosis increased the likelihood of recommending continued hospitalization. We fitted a pooled logistic regression model with the diagnostic-history condition and current-evidence condition as predictors. The psychiatric-background coefficient provided the overall label-effect odds ratio across the five models evaluated. A nested pooled model including the label-by-current-evidence interaction tested whether the diagnostic history contributed a constant additive shift to the log-odds of hospitalization across evidence conditions.

Model-specific analyses compared hospitalization recommendations between diagnostic-history conditions within each model and the current-evidence condition. Absolute risk differences were reported with Newcombe 95% confidence intervals, alongside odds ratios with 95% confidence intervals; risk ratios were provided descriptively. The ten model-by-evidence contrasts were tested using two-sided Fisher's exact tests, with Holm correction for multiple comparisons. Heterogeneity in the label effect was examined using a factorial logistic regression that included psychiatric background, current evidence, model, all two-way interactions, and the three-way interaction. Nested likelihood-ratio tests evaluated the label-by-model interaction and whether this interaction differed between current-evidence conditions.

Confidence was analyzed separately within each model and current-evidence condition. Mean differences were reported with 95% confidence intervals and standardized using Cohen's *d*. Group differences were evaluated using Welch's two-sample *t* tests, with Holm correction across the ten contrasts. We also fitted a linear regression of confidence on psychiatric background, current evidence, and model, using HC3 heteroskedasticity-robust standard errors. A sensitivity analysis was additionally adjusted for the disposition recommendation. Because disposition may mediate the association between psychiatric background and confidence, this model was not interpreted as an estimate of the total label effect.

Reference concordance and signal-detection measures were examined as exploratory outcomes. For the signal-detection analysis, recommending hospitalization when voices persisted was defined as a hit, whereas recommending hospitalization when voices had ceased was defined as a false alarm. These rates were used to calculate sensitivity ($d'$) and decision criterion ($c$).

Repeated calls were treated as stochastic draws from each model's output distribution under a fixed prompt. They do not represent independently sampled patients or clinical cases. Accordingly, the inferential results characterize the specific model-prompt -vignette combinations examined and are not intended to support statistical generalization to other models, prompts, patients, or clinical presentations.

### Ethics, transparency, and reproducibility

The computational experiment used a synthetic case and did not involve patients, medical records, or identifiable patient information. To support reproducibility, we have released the prompt, four vignette versions, processing code, cell-level counts, response-level de-identified outputs where provider terms permit, and analysis code. Because hosted models may change, access dates and endpoint identifiers are essential to interpretation.

## Results

### Analytical sample

The experiment yielded 20,000 classified outputs: 5 models × 2 psychiatric-background conditions × 2 current-evidence conditions × 1,000 repeated calls. The primary analysis compared the proportion recommending continued hospitalization between label conditions within each model and the current-evidence condition (Table 1 and Figure 2).

### Prior schizophrenia diagnosis shifted recommendations toward hospitalization.

Across all five models and in both current-evidence conditions, adding a documented prior diagnosis of schizophrenia increased recommendations for continued hospitalization. In the primary pooled logistic regression, adjusted for current-evidence condition, the prior diagnosis was associated with higher odds of recommending continued hospitalization (OR, 4.48; 95% CI, 3.98–5.03; Figure 2). The label effect differed between current-evidence conditions (likelihood-ratio $\chi^2(1) = 38.70$, $p < .001$), varied across models (label-by-model likelihood-ratio $\chi^2(4) = 60.31$, $p < .001$), and showed further model-specific variation across evidence conditions (three-way likelihood-ratio $\chi^2(4) = 46.43$, $p < .001$). The ratio of the label-associated odds ratios for the voices-persisted versus voices-ceased conditions was 0.455 (95% CI, 0.357–0.581), indicating that the relative label effect was approximately 2.2 times larger when the voices had ceased.

In the discharge-supporting condition, in which the voices had ceased, the hospitalization rate across the five models increased from 4.7% (237/5,000) without a stated prior psychiatric diagnosis to 22.3% (1,113/5,000) with a prior schizophrenia diagnosis. This corresponded to an absolute increase of 17.5 percentage points (95% CI, 16.2–18.8) and a risk ratio of 4.70.

The largest absolute shift occurred in Mistral Large 2512, where hospitalization recommendations increased from 3.3% to 37.7% (RD: 34.4 percentage points). For Grok 4.6, the hospitalization recommendation rate increased from 1.2% to 15.7%. This 14.5-percentage-point increase corresponded to a 13.1-fold relative increase, the largest risk ratio among the five models in this condition. Without the label, Grok 4.6 also had the highest reference concordance in the discharge-supporting condition (98.8%). GPT-5.6 Sol showed the smallest absolute shift, although hospitalization recommendations still increased from 6.4% to 16.8% (RD, 10.4 percentage points; RR, 2.63). Claude Sonnet 5 and Gemini 3.6 Flash showed intermediate absolute increases of 14.5 and 13.8 percentage points, respectively (Table 1).

In the hospitalization-supporting condition, where voices had decreased but remained intermittently present, hospitalization recommendations were already frequent without the diagnostic label. Across the five models, rates increased from 92.6% (4,630/5,000) without the label to 97.0% (4,852/5,000) with it (RD, 4.44 percentage points; 95% CI, 3.58–5.31; RR, 1.05). Model-specific absolute increases ranged from 2.4 to 7.0 percentage points. These smaller absolute shifts occurred against no-label hospitalization rates of 89.4%–94.6%, leaving comparatively little scope for further increases.

The diagnostic-label effect was positive in all ten model-by-evidence contrasts. Each contrast remained statistically significant in two-sided Fisher's exact tests after Holm correction (all adjusted $p \leq .014$). Model-specific counts, proportions, reference-concordance rates, confidence estimates, and diagnostic-label effects are reported in Table 1.

**Label-associated change in reference concordance differed across evidence conditions.**

In the discharge-supporting condition, the shift toward hospitalization necessarily reduced concordance with the prespecified reference disposition. Descriptively pooled reference concordance fell from 95.3% without the label to 77.7% with it. In the hospitalization-supporting condition, by contrast, the same directional shift raised concordance from 92.6% to 97%. This asymmetry is central to interpretation: psychiatric history did not simply make the models less accurate according to the experimental reference. It consistently shifted their decision criterion toward hospitalization, improving concordance in one current-evidence condition while worsening it in the other. This reversal provides the empirical leverage for evaluating a prior-updating account: a common shift toward hospitalization should have opposing consequences for concordance depending on the disposition supported by the current evidence, motivating the interaction analyses across evidence conditions and models. Because the reference dispositions were prespecified by the investigators rather than treated as definitive clinical ground truth, these results are termed reference concordance rather than clinical accuracy.

**Confidence increased even when reference concordance declined.**

Adding the prior schizophrenia diagnosis increased stated confidence in all ten model-by-evidence contrasts. Mean increases ranged from 1.67 to 5.25 points on the 0–100 scale, with Cohen's *d* values ranging from 0.35 to 1.09; all 95% confidence intervals excluded zero, and all contrasts remained statistically significant after Holm correction (all adjusted $p<.001$; Table 1; Figure 3).

This divergence was most pronounced in the discharge-supporting condition. Across models, the same label-induced shift that reduced descriptively pooled reference concordance from 95.26% to 77.74% simultaneously increased confidence. For GPT-5.6 Sol, mean confidence increased from 87.13 to 91.80 (mean difference, 4.67 points; 95% CI, 4.19–5.15; Cohen's $d=0.86$). Corresponding increases were observed for Gemini 3.6 Flash, from 88.96 to 93.20 (mean difference, 4.24; 95% CI, 3.71–4.77; $d=0.70$), and Mistral Large 2512, from 86.41 to 90.57 (mean difference, 4.16; 95% CI, 3.65–4.67; $d=0.71$). Thus, when the current evidence supported discharge, the models became less concordant with the experimental reference and more confident in the same step.

In the response-level linear model adjusted for current evidence and model, the prior diagnosis was associated with a 3.77-point increase in confidence (95% CI, 3.63–3.92; $p<.001$). The association was similar in a sensitivity analysis that additionally adjusted for the disposition recommendation (mean difference, 3.82 points; 95% CI, 3.68–3.97; $p<.001$). Because disposition may lie on the pathway between psychiatric background and confidence, this estimate is presented as a sensitivity analysis rather than interpreted as the direct causal effect of the diagnostic label.

**Exploratory signal-detection analysis**

Exploratory signal-detection analyses treated the two current-evidence conditions as reference states between which the models had to discriminate. Adding the prior schizophrenia diagnosis shifted the decision criterion toward continued hospitalization in every model, with changes in criterion ranging from −0.49 to −1.04; more negative criterion values indicate a greater tendency to recommend continued hospitalization (Figure 2a).

Changes in sensitivity separated the models into two descriptive patterns. For GPT-5.6 Sol, $d'$ changed only modestly, from 3.01 in the no-prior-diagnosis-stated condition to 2.87 in the prior-schizophrenia-diagnosis condition ($\Delta d' = -0.13$). Gemini 3.6 Flash likewise showed essentially unchanged sensitivity ($d'$, 2.74 vs. 2.77; $\Delta d' = +0.03$). In these two models, the diagnostic-label effect was therefore expressed primarily as a shift in decision criterion, with little evidence of reduced discrimination between the two current-evidence conditions (Supplementary Table 2).

By contrast, sensitivity declined more substantially for Claude Sonnet 5 ($d'$, 3.32 to 2.72; $\Delta d' = -0.60$), Grok 4.6 (3.86 to 2.92; $\Delta d' = -0.95$), and Mistral Large 2512 (3.09 to 2.11; $\Delta d' = -0.97$). For these models, the criterion shift toward hospitalization was accompanied by reduced separation between the voices-ceased and voices-persisted conditions (Supplementary Table 2).

These findings support a cross-model shift in response tendency while suggesting that its consequences for discrimination differed across systems. They remain exploratory because the indices were derived from two evidence states within a single reconstructed case and therefore do not estimate discrimination across the broader clinical domain.

**Figure 2. The diagnostic label shifts every model's decision criterion and alters the separation between the two current-evidence conditions.**
(a) Receiver operating characteristic space for each model. Arrows point from responses with no prior diagnosis stated (open circles) to responses with a prior schizophrenia diagnosis (filled circles). False alarms are recommendations to continue hospitalization when the voices had ceased; hits are such recommendations when the voices persisted. Grey curves mark equal-variance iso-d′ contours. Movement along a contour reflects a change in decision criterion; movement across contours reflects a change in sensitivity.
(b) Odds ratios for the effect of the prior diagnosis, shown separately for each model and current-evidence condition (95% confidence intervals). Orange denotes ceased voices, and blue denotes persistent voices. Each estimate is based on 1,000 responses per diagnosis condition (*n* = 2,000). The dashed line marks OR = 1. Differences between the orange and blue estimates within a model indicate that the label's effect varied with the current evidence.

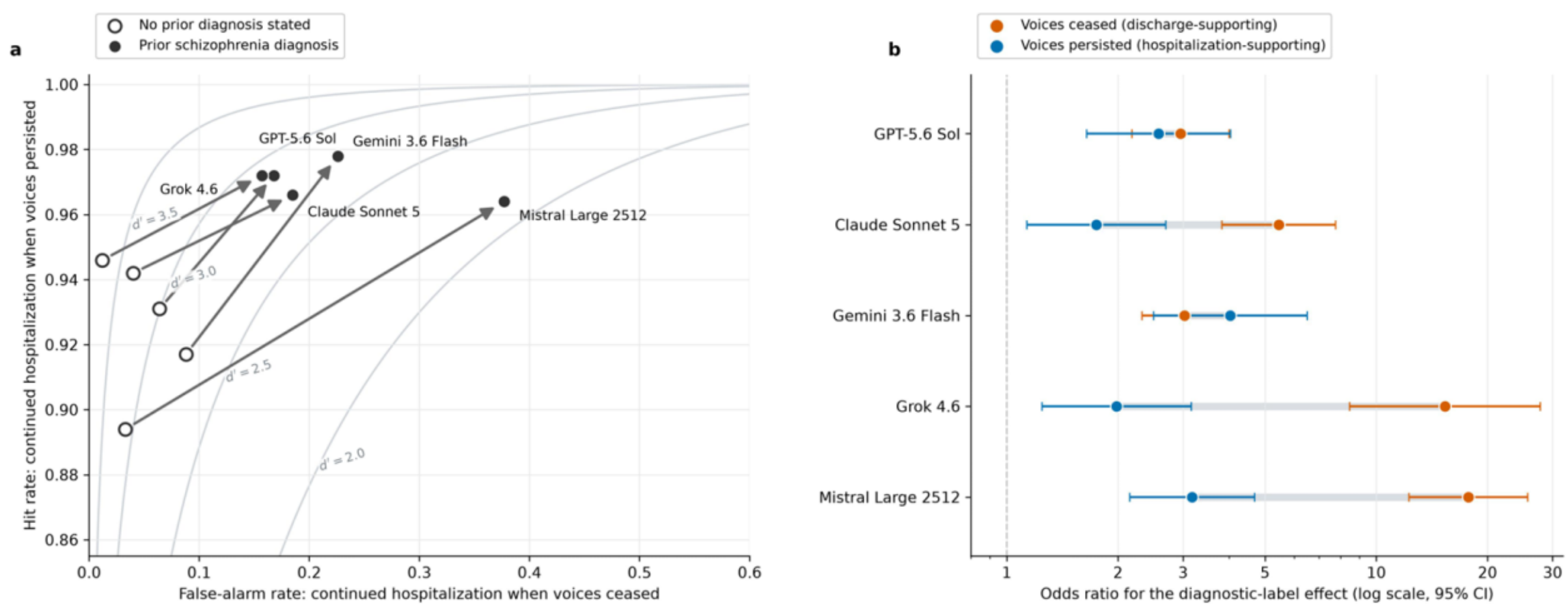

**Figure 3.** Mean stated decision confidence across the four experimental conditions. For each model, orange indicates that the voices had ceased (discharge-supporting evidence), whereas blue indicates that they persisted (hospitalization-supporting evidence). Open circles indicate no prior diagnosis stated, and filled circles indicate a prior schizophrenia diagnosis. Lines connect the two diagnostic-history conditions within each model-by-evidence contrast. Each point represents 1,000 responses, and each connected contrast therefore includes 2,000 responses. Confidence was consistently higher when the voices persisted than when they had ceased, and a prior schizophrenia diagnosis increased confidence in both current-evidence conditions across all five models.

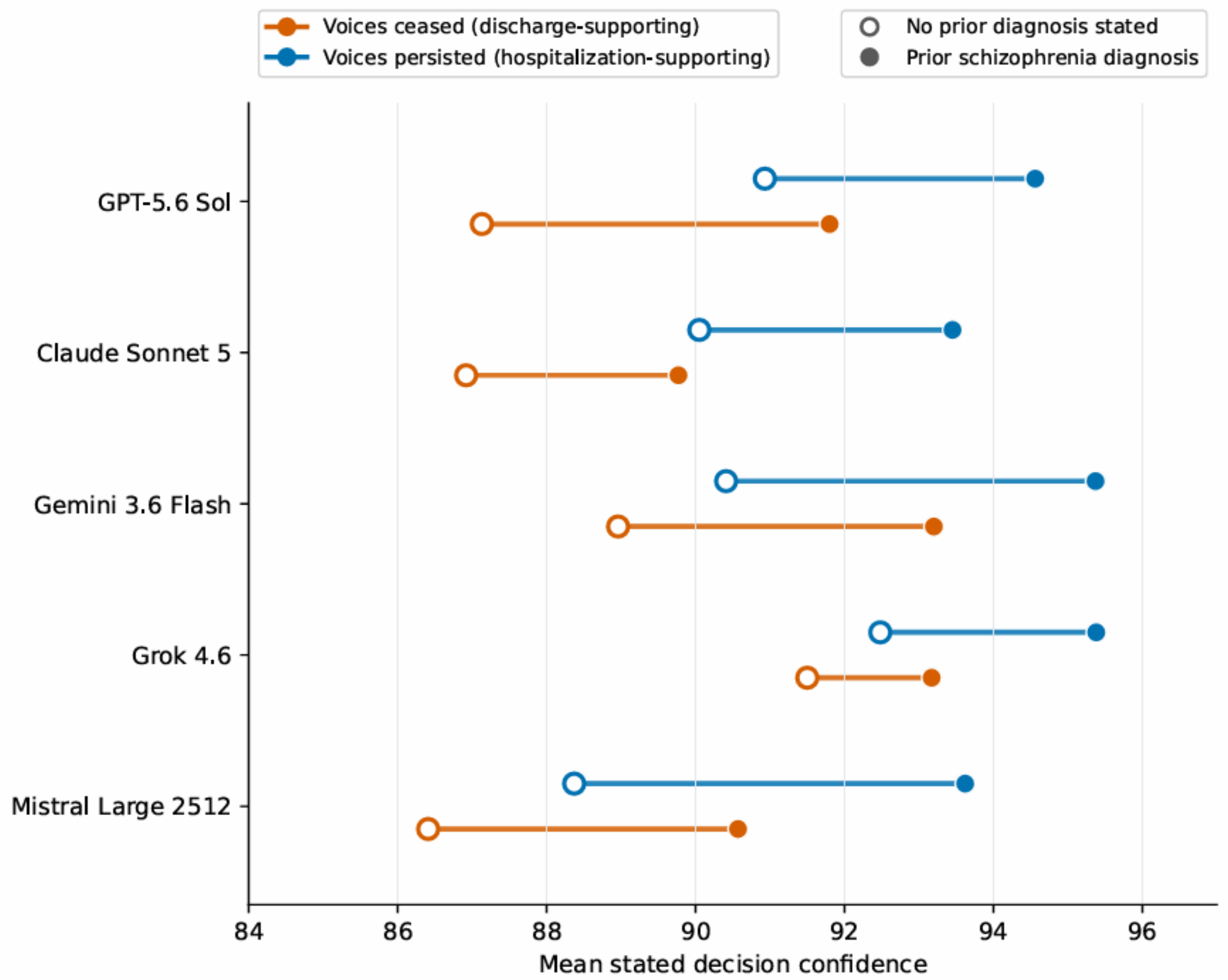

***Table 1.*** *Disposition recommendations, reference concordance, confidence, and diagnostic-label effects by model and current-evidence condition.*

| Model | Current-evidence condition | Prior diagnosis absent: Discharge / hospitalization, n (%) | Prior diagnosis present: Discharge / hospitalization, n (%) | Reference concordance, %: Absent / present | Confidence, mean ± SD: Absent / present | Risk difference, pp (95% CI) | Odds ratio (95% CI) | Fisher's exact *p* | Holm-adjusted *p* |
|---|---|---|---|---|---|---|---|---|---|
| GPT-5.6 Sol | Voices ceased (supports discharge) | 936 (93.6) / 64 (6.4) | 832 (83.2) / 168 (16.8) | 93.6 / 83.2 | 87.13 ± 6.23 / 91.80 ± 4.55 | 10.4 (7.6–13.2) | 2.95 (2.18–4.00) | <.001 | <.001 |
| GPT-5.6 Sol | voices persisted (supports hospitalization) | 69 (6.9) / 931 (93.1) | 28 (2.8) / 972 (97.2) | 93.1 / 97.2 | 90.93 ± 5.41 / 94.56 ± 3.91 | 4.1 (2.2–6.0) | 2.57 (1.64–4.03) | <.001 | <.001 |
| Claude Sonnet 5 | Voices ceased (supports discharge) | 960 (96.0) / 40 (4.0) | 815 (81.5) / 185 (18.5) | 96.0 / 81.5 | 86.92 ± 4.70 / 89.77 ± 6.48 | 14.5 (11.8–17.2) | 5.45 (3.82–7.76) | <.001 | <.001 |
| Claude Sonnet 5 | Voices intermittent (supports hospitalization) | 58 (5.8) / 942 (94.2) | 34 (3.4) / 966 (96.6) | 94.2 / 96.6 | 90.05 ± 5.35 / 93.45 ± 3.82 | 2.4 (0.6–4.3) | 1.75 (1.13–2.70) | .014 | .014 |
| Gemini 3.6 Flash | Voices ceased (supports discharge) | 912 (91.2) / 88 (8.8) | 774 (77.4) / 226 (22.6) | 91.2 / 77.4 | 88.96 ± 7.03 / 93.20 ± 4.94 | 13.8 (10.7–16.9) | 3.03 (2.32–3.94) | <.001 | <.001 |
| Gemini 3.6 Flash | Voices intermittent (supports hospitalization) | 83 (8.3) / 917 (91.7) | 22 (2.2) / 978 (97.8) | 91.7 / 97.8 | 90.41 ± 5.22 / 95.37 ± 3.74 | 6.1 (4.2–8.1) | 4.02 (2.49–6.49) | <.001 | <.001 |
| Grok 4.6 | Voices ceased (supports discharge) | 988 (98.8) / 12 (1.2) | 843 (84.3) / 157 (15.7) | 98.8 / 84.3 | 91.50 ± 4.83 / 93.17 ± 4.71 | 14.5 (12.2–16.9) | 15.33 (8.46–27.78) | <.001 | <.001 |
| Grok 4.6 | Voices intermittent (supports hospitalization) | 54 (5.4) / 946 (94.6) | 28 (2.8) / 972 (97.2) | 94.6 / 97.2 | 92.48 ± 4.09 / 95.38 ± 3.10 | 2.6 (0.9–4.4) | 1.98 (1.24–3.16) | .005 | .009 |
| Mistral Large 2512 | Voices ceased (supports discharge) | 967 (96.7) / 33 (3.3) | 623 (62.3) / 377 (37.7) | 96.7 / 62.3 | 86.41 ± 6.73 / 90.57 ± 4.83 | 34.4 (31.2–37.6) | 17.73 (12.25–25.67) | <.001 | <.001 |
| Mistral Large 2512 | Voices intermittent (supports hospitalization) | 106 (10.6) / 894 (89.4) | 36 (3.6) / 964 (96.4) | 89.4 / 96.4 | 88.37 ± 6.95 / 93.62 ± 4.05 | 7.0 (4.8–9.3) | 3.17 (2.15–4.68) | <.001 | <.001 |

*Note.* Across the four vignette versions, all information other than the current evidence and psychiatric background was held constant. Each of the five models completed 1,000 independent calls per experimental cell. Disposition values are presented as discharge / continued hospitalization. Comparisons contrast the prior-diagnosis-present condition with the prior-diagnosis-absent condition within each model and current-evidence condition. Positive risk differences indicate a higher rate of continued-hospitalization recommendations when a prior schizophrenia diagnosis was documented. Odds ratios greater than 1 indicate higher odds of continued hospitalization in the prior-diagnosis-present condition. Two-sided Fisher's exact tests were used, with Holm correction across all ten comparisons. *pp* = percentage points; CI = confidence interval.

## Discussion

Across five LLMs, a prior diagnosis of schizophrenia shifted recommendations toward continued hospitalization, although the current clinical information was unchanged within each evidence condition. The largest absolute shift occurred when the voices had ceased and the vignette supported discharge: recommendations for continued hospitalization increased from 4.7% without the diagnosis to 22.3% with it. The effect was present in all five models and in both evidence conditions. These findings do not establish that LLMs possess a human psychological bias. They show that information recorded as diagnostic history can alter a model's response to an otherwise identical clinical presentation.

Continued hospitalization served here as a consequential and measurable test of the influence of diagnostic history. The underlying issue, however, is broader. Once a psychiatric diagnosis is entered into a record, it may become a durable frame through which later behavior, testimony, and signs of improvement are interpreted (Langer & Abelson, 1974; Sims et al., 2021). In psychiatry, where clinical states often change over time, a description of the past can gradually come to stand for the person in the present. The concern is not simply that an earlier diagnosis affects one decision, but that it may continue to organize judgment even after the presentation it originally described has changed (Altmann et al., 2024; Yanos et al., 2010).

These findings echo the central epistemic question raised by Rosenhan (Rosenhan, 1973). His study has been extensively criticized on methodological, conceptual, and historical grounds, and our findings should not be read as validating its empirical claims (Cahalan, 2019; Scull, 2023; Spitzer, 1975). Yet the epistemic problem it dramatized remains important: once a psychiatric diagnosis becomes part of the record, can observers remain responsive to evidence that the person's current state has changed? This question has particular force in psychiatry, where symptoms, functioning, and risk can fluctuate over time, while diagnostic categories may persist and continue to organize interpretation (Palomar-Ciria et al., 2019). For clinicians, holding history in mind is indispensable, but so is the capacity to revise what that history means in light of the present. AI systems sharpen this tension. Their ability to retain and integrate information across long records may exceed ordinary human memory, but that same capacity can make historical formulations unusually persistent. What appears to be an advantage, the ability to remember more may therefore also become a source of risk when the system fails to recognize that the patient has changed.

It is not to say that psychiatric history is irrelevant or inherently stigmatizing. A diagnosis of schizophrenia may carry clinically useful information about illness course, recurrence, and future risk; previous relapses and hospitalizations, for example, are associated with subsequent relapse risk (Olivares et al., 2013). In Bayesian terms, such history may appropriately alter the prior probability assigned to the continuation of illness (Bours, 2021). The question is therefore not whether a model should use diagnostic history, but how the weight of that history should change as new, person-specific evidence becomes available.
A simple additive account would predict that the diagnosis contributes a constant increment to the log-odds of continued hospitalization, yielding the same label-associated odds ratio in both current-evidence conditions. That was not the observed pattern. The label effect differed significantly according to whether the voices had ceased or persisted, and was approximately 2.2 times larger when the voices had ceased. The diagnosis, therefore, exerted its largest relative effect when the current presentation most clearly departed from the historical diagnosis. A uniform shift in prior probability does not fully account for this result, though it remains consistent with more complex forms of evidence integration or decision-making.
The high baseline hospitalization rate in the voices-persisted condition warrants caution. Even without the diagnosis, 92.6% of responses recommended continued hospitalization, leaving little scope for a further absolute increase. This ceiling readily explains part of the asymmetry on the probability scale. It does not, by itself, account for the interaction on the log-odds scale, because a constant addition to the log-odds would yield the same odds ratio despite different baseline probabilities. Nevertheless, odds ratios at extreme event rates can be sensitive to small changes in the number of discordant responses. The interaction argues against the simplest constant-additive account, but it cannot establish the process that generated the difference.

The signal-detection analysis provides a complementary description of this pattern. Adding the diagnosis shifted the decision criterion toward continued hospitalization in every model; because lower values of $c$ indicate a greater tendency to recommend hospitalization, the negative changes in $c$ reflect a consistent movement toward that response. These model-specific patterns remain exploratory, however, because they were derived from only two versions of a single clinical scenario. Confidence moved in the same direction as the disposition recommendations. Adding the diagnosis increased stated confidence in all ten model-by-evidence contrasts. In the discharge-supporting condition, confidence rose while pooled concordance with the prespecified experimental reference fell from 95.3% to 77.7%. The models thus became less concordant with the reference while expressing their recommendations more confidently. Stated confidence should not be interpreted as a calibrated probability, still less as evidence that a model experiences certainty. It nevertheless captures how strongly the recommendation was expressed.

The experiment cannot determine why the models produced this pattern. Training data may encode associations between schizophrenia, chronic illness, risk, and prolonged hospitalization. Post-training procedures may also favor avoiding premature discharge over avoiding unnecessary continued hospitalization when uncertainty exists. Previous work has shown that model state and post-training configuration can alter both model outputs and bias-related behavior (Ben-Zion et al., 2025; Coda-Forno et al., 2023). These possibilities are not mutually exclusive. Because the models were proprietary and the experiment manipulated their inputs rather than their internal states, the findings cannot be attributed to Reinforcement Learning from Human Feedback or to any other particular mechanism.

Concepts drawn from human clinical reasoning require similar restraint. The pattern resembles anchoring or diagnostic overshadowing, but resemblance in output does not imply a shared cognitive process. Therefore, we use *label sensitivity* descriptively: the addition of a prior diagnosis changed the distribution of recommendations despite identical current clinical information. Nor should label sensitivity be equated automatically with error. Some influence of diagnostic history may be justified by its prognostic value. The potential problem lies in the transition from a group-level association, where people with a particular history may, on average, face a higher risk, to a conclusion about the person currently being assessed. A statistical prior is useful only if it remains open to revision by individual evidence.

Our study has several limitations. The systems evaluated were general-purpose commercial models, not tools developed or validated for psychiatric disposition decisions, and performance in a controlled experiment cannot establish their safety in clinical practice (Hager et al., 2024). The study used four versions of one synthetic vignette and a single standardized prompt. This design isolated the diagnostic-history and current-evidence manipulations, but it does not show whether the same pattern would occur with other diagnoses, presentations, prompts, or decisions. The vignette described a White man, consistent with the Rosenhan-inspired scenario, and the results may not generalize across sociodemographic groups. Calls were made via remotely hosted APIs over eight days, and unannounced provider-side changes during that period cannot be ruled out. Repeated calls captured variability under fixed experimental conditions; they did not represent independent patients or clinical cases. Finally, the reference dispositions were prespecified experimental categories rather than definitive clinical ground truth.

The findings nevertheless suggest a concrete direction for evaluating psychiatric AI. Systems that process longitudinal records should be assessed not only on whether they retain relevant history but also on whether they revise the meaning and weight of that history as symptoms, functioning, and circumstances change. Counterfactual comparisons such as those used here can reveal whether the same current presentation is interpreted differently when a historical diagnosis is added or removed. Establishing whether such a difference is clinically appropriate will require richer longitudinal cases, outcome data, and comparison with expert judgment. The aim is not invariance to history, but responsiveness to change.

The same distinction matters when clinicians interpret model outputs. A recommendation may reflect a statistically valid group-level association without adequately characterizing the individual patient. It should therefore be possible to determine which elements of the response arise from the diagnostic history and which arise from the current presentation. In this limited sense, psychiatric AI may require something analogous to clinical supervision: recurrent scrutiny of how prior formulations direct attention, what evidence they bring into view, and what evidence they obscure. For an AI system, this could be operationalized through structured justification, counterfactual evaluation, and explicit separation of historical from current evidence.

These safeguards should be embedded within human oversight rather than treated as substitutes for clinical judgment. When a model recommends continued hospitalization despite current evidence of improvement, the discrepancy could be designed to prompt explicit clinician review and documentation of the current risk findings that support the restrictive recommendation. Institutions should also monitor label sensitivity prospectively across diagnoses, demographic groups, model versions, and clinical settings. This approach is consistent with guidance emphasizing transparency, task-specific evaluation, human control, and ongoing governance in clinical AI (Gallifant et al., 2025; Hager et al., 2024; Mosidze et al., 2026). The goal is not to remove psychiatric history from decision-making, but to ensure that its influence remains visible, reviewable, and responsive to changes in the individual patient.

In this experiment, the diagnosis shifted a consequential minority of decisions, but it did not generally displace the current evidence. Even with the diagnosis present, 77.7% of responses in the discharge-supporting condition still recommended discharge. The findings, therefore, do not show that psychiatric history is harmful or that models should ignore it. They reveal a problem of dynamic integration: how to preserve the predictive value of what came before while allowing evidence about the individual patient to modify its influence. A clinically useful system must not only remember the patient's history; it must also recognize when the patient has changed.

## Declarations

### Ethics statement

The computational experiment used entirely synthetic clinical vignettes and did not involve patients, medical records, or identifiable patient information.

### Funding

This research received no specific grant from any funding agency in the public, commercial, or not-for-profit sectors.